\documentclass[twocolumn]{aa}

\usepackage{amsmath}
\usepackage[varg]{txfonts}

\begin{document}

\title{Probing solar flare accelerated electron distributions with prospective X-ray polarimetry missions}

\author{Natasha L. S. Jeffrey\inst{\ref{inst1}}
\and Pascal Saint-Hilaire\inst{\ref{inst2}}
\and
Eduard P. Kontar\inst{\ref{inst3}}}
\institute{Department of Mathematics, Physics \& Electrical Engineering, Northumbria University, Newcastle upon Tyne, UK, NE1 8ST, \email{natasha.jeffrey@northumbria.ac.uk}\label{inst1}
\and
Space Science Laboratory, University of California, Berkeley, USA\label{inst2}
\and
School of Physics \& Astronomy, University of Glasgow, Glasgow, UK, G12 8QQ\label{inst3}
}

\date{Received \today /Accepted }

\titlerunning{Probing flare accelerated electrons with X-ray polarization}
\authorrunning{Jeffrey et al.}

\abstract{
Solar flare electron acceleration is an extremely efficient process, but the method of acceleration is not well constrained. Two of the essential diagnostics, electron anisotropy (velocity angle to the guiding magnetic field) and the high energy cutoff (highest energy electrons produced by the acceleration conditions: mechanism, spatial extent, and time), are important quantities that can help to constrain electron acceleration at the Sun but both are poorly determined. Here, by using electron and X-ray transport simulations that account for both collisional and non-collisional transport processes, such as turbulent scattering and X-ray albedo, we show that X-ray polarization can be used to constrain the anisotropy of the accelerated electron distribution and the most energetic accelerated electrons together. Moreover, we show that prospective missions, for example CubeSat missions without imaging information, can be used alongside such simulations to determine these parameters. We conclude that a fuller understanding of flare acceleration processes will come from missions capable of both X-ray flux and polarization spectral measurements together. Although imaging polarimetry is highly desired, we demonstrate that spectro-polarimeters without imaging can also provide strong constraints on electron anisotropy and the high energy cutoff.
}


\maketitle

\section{Introduction}\label{intro}
Solar flares are the observational product of magnetic energy release in the Sun's atmosphere and the conversion of this magnetic energy into kinetic energies. Processes of energy release and transfer acting in flares are initiated by magnetic reconnection in the corona \citep[e.g.][]{1957JGR....62..509P,1958IAUS....6..123S,2000mare.book.....P}. Many studies show that a substantial fraction of the magnetic energy goes into the acceleration of energetic keV and sometimes MeV electrons \citep[e.g.][]{2012ApJ...759...71E,2015ApJ...802...53A,2017ApJ...836...17A,2016A&A...588A.116W}. Although it is established that solar flares are exceptionally efficient particle accelerators and hence, a relatively close astrophysical laboratory for studying particle acceleration, the exact mechanisms and even the exact locations of energy release and/or acceleration are not well-constrained \citep[e.g.][]{2017LRSP...14....2B}. The magnetic energy may be dissipated to particles by plasma turbulence \citep[e.g.][]{1993ApJ...418..912L,2012SSRv..173..535P,2016ApJ...827L...3V,2017PhRvL.118o5101K}, but it is possible that other mechanisms, such as shock acceleration in reconnection outflows, contribute as well \citep[e.g.][]{2009A&A...494..669M,2015Sci...350.1238C}. Further, the acceleration of seemingly distinct populations of electrons at the Sun and those detected in-situ in the heliosphere, as well as the connection between flare-accelerated electrons and MeV to GeV ions, is poorly understood.

X-ray bremsstrahlung is the prime diagnostic of flare-accelerated electrons at the Sun \citep[e.g.][]{2011SSRv..159..301K}. However, one reason that the properties of the acceleration region remain exclusive is because bremsstrahlung emission is density weighted, and so flare-accelerated electrons produce their strongest emission away from the primary site(s) of acceleration in the corona, in the chromosphere where the density is high. In a standard model, electrons accelerated in the corona stream towards the dense layers of the chromosphere where they lose energy via collisions and produce hard X-ray (HXR) footpoints \citep[e.g.][]{2011SSRv..159..107H}. Hence, in order to determine the properties of accelerated electrons and the environment in which they are accelerated, we require state-of-the art transport models and diagnostic tools that realistically account for collisions in both the corona and chromosphere, and constrain non-collisional transport effects such as turbulent scattering \citep{2014ApJ...780..176K}. In the last few years, there has been significant advancement in our understanding of electron transport at the Sun. Instead of modelling electron transport with very little interaction with the coronal plasma, we now understand the importance of accounting for diffusive processes in energy and pitch-angle in the corona \citep[e.g.][]{2014ApJ...787...86J, 2015ApJ...809...35K}. For example, the application of a full collisional model, accounting for energy diffusion, led to a solution of the `low-energy cutoff' problem \citep[e.g. ][]{2019ApJ...871..225K}, whereby the energy associated with flare-accelerated electrons could not be constrained from the X-ray flux spectrum. However, many of the vital properties required to constrain the acceleration process(es) still remain elusive since they are difficult to determine from a single X-ray flux spectrum alone.

Since the birth of X-ray astronomy, X-ray polarization in solar physics has been understudied. Observations and studies exist, but many results have large uncertainties \citep[e.g.][]{2011SSRv..159..301K}. This is mainly because the polarimeters were unsuitable for flare observations; they were secondary add-on missions, such as the polarimeter on board the Reuven Ramaty High Energy Solar Spectroscopic Imager (RHESSI, \citet{2002SoPh..210....3L,2002SoPh..210..125M}), or they were not optimised for solar observations e.g. they were astrophysical missions studying gamma ray bursts. However, the X-ray polarization spectrum is an observable that can provide a direct link to several key properties of energetic electrons, including the anisotropy, which is usually an unknown quantity that is vital for constraining both acceleration and transport properties in the corona. Many studies have extensively modelled both spatially integrated and spatially resolved solar flare X-ray polarization, i.e. \citet{1978ApJ...219..705B}, \citet{1983ApJ...269..715L}, \citet{2008ApJ...674..570E}, \citet{2011A&A...536A..93J}. Here, the aim is not to reiterative the main results of these past studies but to show how prospective missions, such as relatively cheap CubeSat missions, can be used alongside electron and X-ray transport simulations to determine vital acceleration parameters, even without imaging information. 

Section \ref{model} describes the electron and X-ray transport models used in this study, while Section \ref{instruments} briefly describes some proposed X-ray spectro-polarimeters. In Section \ref{results}, we show simulation examples of spatially integrated X-ray polarization and demonstrate how acceleration parameters can be determined from spatially integrated X-ray flux and polarization spectra together. We briefly summarise the study in Section \ref{conclusions}.

\section{Electron and X-ray transport}\label{model}

\subsection{Electron transport model}
To determine how the properties of flare-accelerated electrons are changed in a hot and collisional flaring coronal plasma, 
we use the kinetic transport simulation first discussed in \citet{2014ApJ...787...86J} and \citet{2015ApJ...809...35K}. 
We model the evolution of an electron flux 
$F(z,E,\mu)$~[electron erg$^{-1}$ s$^{-1}$ cm$^{-2}$] in space $z$~[cm], energy $E$~[erg], and pitch-angle $\mu$ to a guiding magnetic field, 
using the Fokker-Planck equation of the form \citep[e.g.][]{1981phki.book.....L,1986CoPhR...4..183K}:

\begin{equation}\label{eq: fp_e}
\begin{split}
\mu \frac{\partial F}{\partial z} &= \Gamma m_{e}^2 \;\frac{\partial}{\partial E} \left[ G (u[E] ) \frac{\partial F}{\partial E} + \frac{G (u[E] )}{E} \left ( \frac{E}{k_B T}-1 \right )F \right] \\
&+ \frac{\Gamma m_{e}^2}{4E^2} \; \frac{\partial}{\partial \mu} \left [ (1-\mu^{2}) \biggl ( {\rm erf} (u[E] ) - G (u[E] ) \biggr ) \frac{\partial F}{\partial \mu} \right ]  \\
& + S(E,z,\mu),\,\,\,\,
\end{split}
\end{equation}

where $\Gamma=4\pi e^{4} \ln\Lambda \, n /m_{e}^{2}=2Kn/m_{e}^{2}$, 
and $e$ [esu] is the electron charge, 
$n$ is the plasma number density [cm$^{-3}$] (a hydrogen plasma is assumed), $m_{e}$ is the electron rest mass [g], and $\ln\Lambda$ is the Coulomb logarithm. The variable $u(E)=\sqrt{E/k_B T}$, where $k_{B}$~[erg K$^{-1}$] is the Boltzmann constant and $T$~[K] is the background plasma temperature. $S(E,z,\mu)$ plays the role of the electron flux source function. The first term on the right-hand side of Equation \ref{eq: fp_e} describes energy evolution due to collisions (advective and diffusive terms), while the second term describes the pitch-angle evolution due to collisions.

The functions ${\rm erf}(u)$ (the error function) and $G(u)$ are given by,

\begin{equation}\label{eq:gcha}
{\rm erf}(u)\equiv (2/\sqrt{\pi})\int\limits_{0}^{u}\exp(-t^2) \, dt
\end{equation}
and
\begin{equation}
G(u)=\frac{{\rm erf}(u)-u \, {\rm erf}^{'}(u)}{2u^{2}} \,\,\, .
\end{equation}
Further information regarding these functions and Equation \ref{eq: fp_e} can be found in \citet{2014PhDT........88J}. The error function and G(u) control the lower-energy ($E\approx k_{B}T$) electron interactions ensuring that they become indistinguishable from the background thermal plasma.

Equation (\ref{eq: fp_e}) is a time-independent equation useful for studying solar flares where the electron transport time from the corona to the lower atmosphere is usually shorter than the observational time (i.e. most X-ray spectral observations have integration times of tens of seconds to minutes), but temporal information can be extracted \citep{2019ApJ...880..136J}. 

Equation (\ref{eq: fp_e}) models electron-electron energy losses,
the dominant electron energy loss mechanism in the flaring plasma, and both electron-electron and electron-proton interactions for collisional pitch-angle scattering\footnote{For this the pitch-angle term in Equation (\ref{eq: fp_e}) is multiplied by 2, compared to \citet{2014ApJ...787...86J} and \citet{2019ApJ...880..136J}.}. Equation (\ref{eq: fp_e}) can be easily generalised to model any particle-particle collisions.

The z coordinate traces the dominant magnetic field direction. For simplicity here, in each simulation we use a homogenous coronal number density and temperature. At the boundary with the chromosphere, the number density is set at $n=1\times10^{12}$~cm$^{-3}$ but the number density rises to photospheric densities of $n\approx10^{17}$~cm$^{-3}$ over $\approx3\arcsec$ using the exponential density function shown \citet{2019ApJ...880..136J}. At the chromospheric boundary, the temperature is set at $T\approx0$~MK so that the electron transport model becomes a cold target model \citep{1971SoPh...18..489B} where all $E>>k_{B}T$. The simulation ends when all of the electrons reach the chromosphere and all electrons have been removed when their energy $E=0$~keV. We assume a homogenous magnetic field. Chromospheric magnetic mirroring is not modelled here as it will not change the main results shown in Section \ref{results}, although we plan to include it in future studies.

\begin{figure*}
\centering
\includegraphics[width=0.99\linewidth]{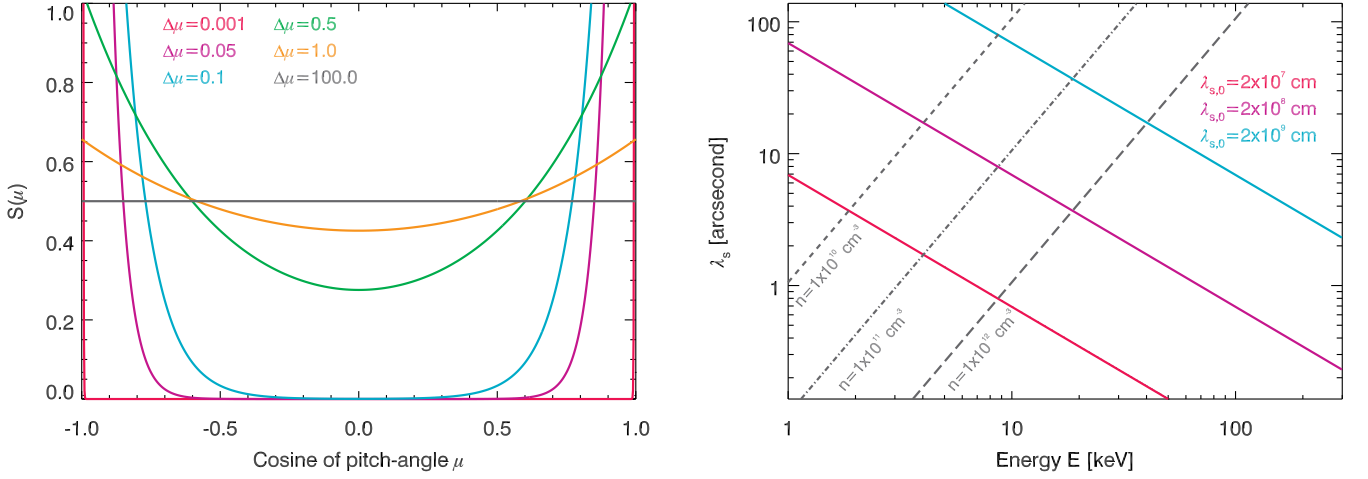}
\caption{Left: Different injected electron anisotropy at the loop apex using $S(\mu)$ (Equation (\ref{eq: mu})). Small $\Delta \mu$ produces beamed distributions while large $\Delta \mu$ produces isotropic distributions. Right: Turbulent scattering mean free path $\lambda_{s}$ versus electron energy $E$ using Equation (\ref{eq:ts}) and using $\lambda_{s,0}=2\times10^{7},\,2\times10^{8}$ and $2\times10^{9}$ cm. Turbulent scattering quickly isotropises higher energy electrons. The mean free path $\lambda_{s}$ is also compared with the collisional mean free path (using $\lambda=v^{4}/\Gamma$; grey dashed and dotted lines) for three different densities of $n=1\times10^{10}$~cm$^{-3}$, $n=1\times10^{11}$~cm$^{-3}$ and $n=1\times10^{12}$~cm$^{-3}$.}
\label{fig0}
\end{figure*}

\subsection{Non-collisional scattering}

We also study how non-collisional transport effects such as turbulent scattering change the electron distribution and the resulting X-ray polarization. Other non-collisional effects can change the electron properties, such as beam-driven Langmuir wave turbulence \citep{2009ApJ...707L..45H}, electron re-acceleration \citep{2009A&A...508..993B} and/or beam-driven return current \citep{1977ApJ...218..306K,1980ApJ...235.1055E,2006ApJ...651..553Z,2017ApJ...851...78A}, but studying these processes is beyond the scope of the paper.

Here we use an isotropic turbulent scattering approximation\footnote{We use this model for turbulent scattering since the details of scattering in the flaring corona are not well-constrained, i.e. there are many models but few observations.} \citep[e.g.][]{1989ApJ...336..243S} where the turbulent scattering diffusion coefficient $D_{\mu\mu}^{T}$ is related to the turbulent scattering mean free path $\lambda_{s}$ [cm] 
and electron velocity $v$ [cm s$^{-1}$] using

\begin{equation}
D_{\mu\mu}^{T}= \frac{v}{2\lambda_{s}}\left(1-\mu^{2}\right).
\end{equation}

Using the standard quasilinear theory for slab turbulence \citep{1966ApJ...146..480J,1966JGR....71....1K,1975MNRAS.172..557S,1982JGR....87.5063L,2014ApJ...780..176K}, $\lambda_{s}$ can be related to the level of turbulent magnetic field fluctuations $\frac{\delta B}{B}$ resonant with electrons of velocity $v$ by

\begin{equation}\label{eq: dB_B}
\lambda_{s}=\frac{v}{\pi \Omega_{ce}}\left(\left<\frac{\delta B^{2}}{B^{2}}\right>\right)^{-1}
=c\frac{\sqrt{2E m}}{\pi e B}\left(\left<\frac{\delta B^{2}}{B^{2}}\right>\right)^{-1},
\end{equation}

where $\Omega_{ce}=eB/m c$ is the electron gyrofrequency [Hz], $B$ is the magnetic field strength [G] and $c$ is the speed of light [cm s$^{-1}$]. Using for example, $E=25$~keV and $B=300$~G, $2\times10^{-4}\le\frac{\delta B}{B}\le2\times10^{-5}$
corresponds to $2\times10^{7}\;[\rm cm]\le \lambda_{s} \le 2\times10^{9}\;[\rm cm]$.

In simulations where we investigate the role of non-collisional turbulent scattering, 
the governing Fokker-Planck equation becomes

\begin{equation}\label{eq: fp_e_new}
\begin{split}
\mu \frac{\partial F}{\partial z} &= \Gamma m_{e}^2 \;\frac{\partial}{\partial E} \left[ G (u[E] ) \frac{\partial F}{\partial E} + \frac{G (u[E] )}{E} \left ( \frac{E}{k_B T}-1 \right )F \right] \\
&+ \frac{\Gamma m_{e}^2}{4E^2} \; \frac{\partial}{\partial \mu} \left [ (1-\mu^{2}) \biggl ( {\rm erf} (u[E] ) - G (u[E] ) \biggr ) \frac{\partial F}{\partial \mu} \right ]  \\
& + \frac{1}{2\lambda_{s}(E)}\frac{\partial}{\partial \mu}\left [(1-\mu^{2}) \frac{\partial F}{\partial \mu} \right] + S(E,z,\mu).\,\,\,\,
\end{split}
\end{equation}

For the majority of coronal flare conditions and electron energies, non-collisional turbulent scattering operates on timescales shorter than collisional scattering and can produce greater isotropy and trapping amongst higher energy electrons (see Figure \ref{fig0}, right panel). By combining X-ray imaging spectroscopy and radio observations of the gyrosynchrotron radiation, \citet{2018A&A...610A...6M} find empirically that
the scattering mean free path is
\begin{equation}\label{eq:ts}
\lambda_{s}=\lambda_{s,0} {\rm [cm]} \left(\frac{25}{E {\rm [keV]}}\right),
\end{equation}
where $\lambda_{s,0}=2\times10^{8}$~cm.  In Section \ref{results}, we use $\lambda_{s,0}=2\times10^{8}$~cm and $\lambda_{s,0}=2\times10^{9}$~cm.

In this model higher energy electrons have a smaller turbulent mean free path than lower energy electrons. The model of \citet{2018A&A...610A...6M} is suitable for the purposes of the paper and while it is based upon the observation of a single flare and has large uncertainties, it clearly shows that the mean free path of higher energy electrons (from microwave observations) is smaller than the mean free path of lower energy electrons (from X-ray observations).

Following \citet{2014ApJ...787...86J}, and re-writting Equation (\ref{eq: fp_e_new}) 
as a Kolmogorov forward equation \citep{Kolmogorov1931}, 
Equation (\ref{eq: fp_e_new}) can be converted to a set of time-independent 
stochastic differential equations (SDEs) 
\citep[e.g. ][]{1986ApOpt..25.3145G,2017SSRv..212..151S} 
that describe the evolution of $z$, $E$, and $\mu$ in It$\hat{\text{o}}$ calculus:

\begin{equation}\label{eq:sto_x}
z_{j+1}=z_{j}+\mu_{j} \, \Delta s \,\,\, ;
\end{equation}
\begin{equation}\label{eq:sto_E}
\begin{split}
E_{j+1} & =E_{j}-\frac{\Gamma m_{e}^{2}}{2E_{j}} \, \bigg ( {\rm erf}(u_{j})-2u_{j}{\rm erf^{\prime}}(u_{j}) \bigg) \, \Delta s\\
& +\sqrt{2 \, \Gamma m_{e}^{2} \, G(u_{j}) \, \Delta s} \,\, W_E \,\,\, ; \,\,\,
\end{split}
\end{equation}
\begin{equation}\label{eq:sto_mu}
\begin{split}
 \mu_{j+1}&=\mu_{j}-\left[\frac{\Gamma m_{e}^{2} \biggl ( {\rm erf}(u_{j})-G(u_{j}) \biggr ) } {2 E_{j}^{2}} \,\, \mu_{j} + \frac{\mu_{j}}{\lambda_{s}(E_{j})}\right] \, \Delta s \\
 &+\sqrt{\left[\frac{ (1-\mu_{j}^{2}) \, \Gamma m_{e}^{2} \, \biggl ( {\rm erf}(u_{j})-G(u_{j}) \biggr ) } {2 E_{j}^{2}} + \frac{\left(1-\mu_{j}^{2}\right)}{\lambda_{s}(E_{j})}\right]\, \Delta s} \, \, W_\mu \,\,\, .
\end{split}
\end{equation}

We note that $\Delta s$~[cm] is the step size along the particle path, 
and $W_\mu$, $W_E$ are random numbers drawn from Gaussian 
distributions with zero mean and a unit variance representing the corresponding Wiener processes \citep[e.g. ][]{1986ApOpt..25.3145G}. A simulation step size of $\Delta s=10^{5}$~cm is used in all simulations, and $E$, $\mu$ and $z$ are updated at each step $j$. A step size of $\Delta s=10^{5}$~cm is approximately two orders of magnitude smaller than the thermal collisional length in a dense ($n=10^{11}$~cm$^{-3}$) plasma with $T\ge 10$~MK (or the collisional length of an electron with an energy of $1$~keV or greater, in a cold plasma). The derivation of Equation (\ref{eq: fp_e_new}) and a detailed 
description of the simulations can be found in \citet{2014ApJ...787...86J}. 

Equation (\ref{eq: fp_e_new}) (and Equations (\ref{eq:sto_E}) and (\ref{eq:sto_mu})) diverge as $E\rightarrow0$, and as discussed in \citet{2014ApJ...787...86J}, 
the deterministic equation $E_{j+1}=\left[E_{j}^{3/2} + \frac{3\Gamma m_{e}^{2}}{2\sqrt{\pi k_{B}T}}\Delta s \right]^{2/3}$ must be used 
for low energies where $E_{j}\le E_{\rm low}$ using $E_{\rm low}=\left[\frac{3\Gamma m_{e}^{2}}{2\sqrt{\pi k_{B} T}} \Delta s\right]^{2/3}$ -- see \citet{2014ApJ...787...86J}, following \citet{2009JCoPh.228.1391L}. For such low energy thermal electrons, $\mu_{j+1}$ can be drawn from an isotropic distribution $\mu \in [-1,+1]$.

Once the electron transport simulations are finished, we also include an additional background coronal thermal component with temperature $T$ and a chosen $EM=n^{2}V$ that is dominant at lower X-ray energies between $\approx1-25$ keV, and where $V$ is the volume of this source. Although it is possible that the thermal component can produce a small detectable polarization of a few percent \citep{1980ApJ...237.1015E}, we assume that the coronal Maxwellian source is isotropic and hence, produces completely unpolarized X-ray emission in all the simulations shown here.

\subsection{Electron input anisotropy and other injection properties}
The initial electron anisotropy is chosen using
\begin{equation}\label{eq: mu}
S(\mu)\propto\frac{1}{2}\exp\left(-\frac{\left(1-\mu \right)}{\Delta\mu}\right)+\frac{1}{2}\exp\left(-\frac{\left(1+\mu \right)}{\Delta\mu}\right)
\end{equation}

where $\Delta\mu$ controls the electron directivity. As $\Delta\mu\rightarrow0$ the distribution is completely beamed, with half directed along one loop leg ($\mu=-1$) and half along the other ($\mu=+1$), and as $\Delta\mu\rightarrow\infty$, the electron distribution becomes isotropic (see the left panel of Figure \ref{fig0}).

In many of the simulation runs shown in Section \ref{results}, the electron directivity is beamed. Since the directivity is unknown, a beamed distribution is used so that differences in the studied parameters are clearly seen and understood.

For most of the simulations shown here, we input sensible flaring parameters: a simple power law distribution in energy ($E^{-\delta}$) with a spectral index of $\delta=5$, a low energy cutoff of $E_{c}=20$~keV and an acceleration rate of $\dot{N}=7\times10^{35}$~electrons s$^{-1}$ and in space, we input a Gaussian at the loop apex with a standard deviation of $1\arcsec$.

\begin{figure*}
\centering
\includegraphics[width=0.59\textwidth]{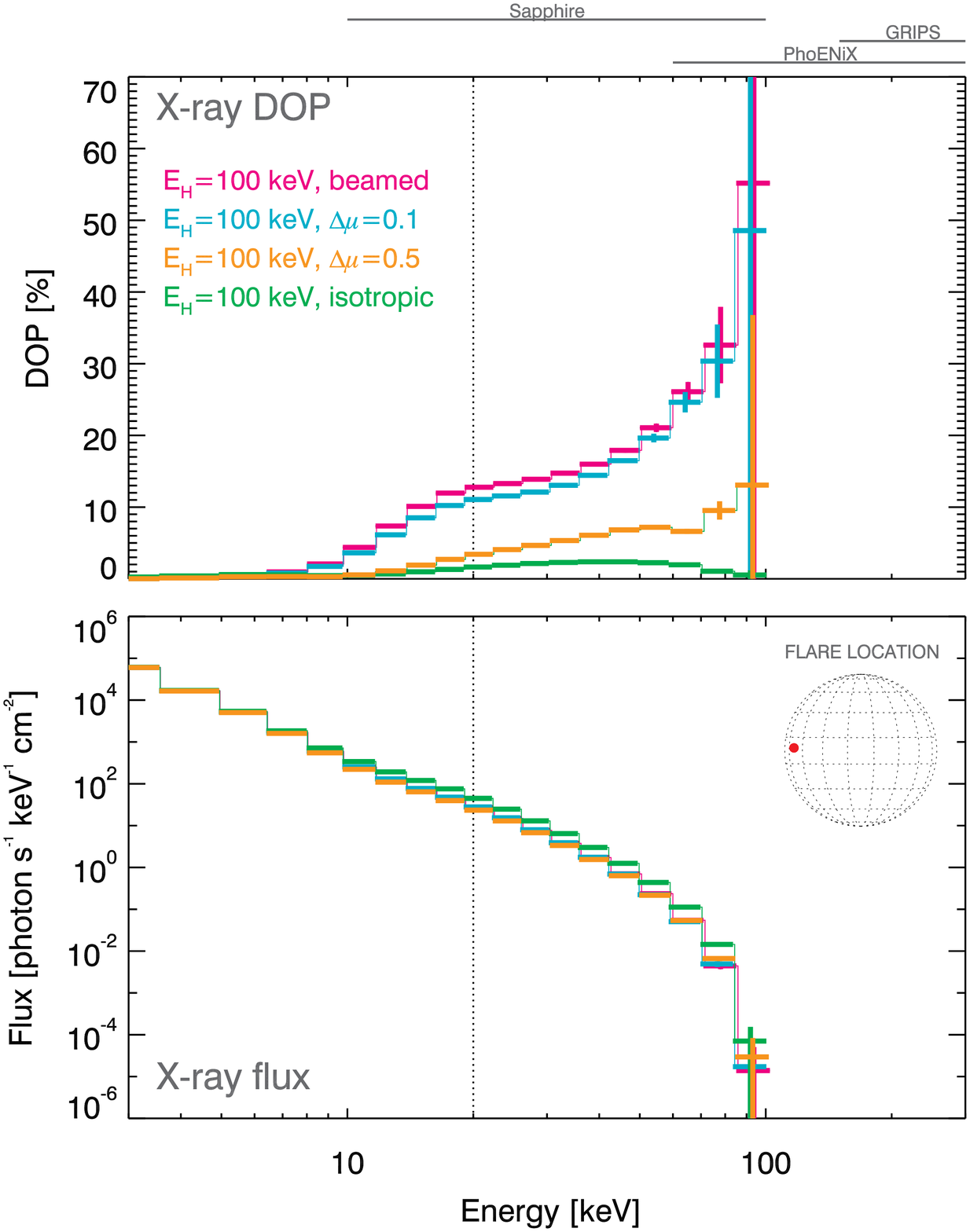}
\caption{Resulting spatially integrated X-ray DOP and flux spectra for three injected electron distributions with either beamed, $\Delta\mu=0.5$, $\Delta\mu=0.1$ or isotropic pitch-angle distributions and using the following identical electron properties of: $\delta=5$, $E_{c}=20$~keV (vertical grey dotted line), $E_{H}=100$~keV and $\dot{N}=7\times10^{35}$~e $s^{-1}$, and corona plasma properties of: $n=3\times10^{10}$~cm$^{-3}$ and $T=20$~MK, plotted for a flare located at a heliocentric angle of $\theta=60^{\circ}$. All spectra include an albedo component and a coronal background thermal component with $EM=n^{2}V=0.9\times10^{48}$~cm$^{-3}$, using a chosen $V=10^{27}$~cm$^{3}$. For example error calculations, we use an effective area of $5$~cm$^{2}$ and a time bin of $120$~s. No turbulent scattering is present.}  \label{fig1}
\end{figure*}

\subsection{Creation of the X-ray distribution}
The electron flux spectrum is calculated and it is converted to a photon spectrum using the full angle-dependent polarization bremsstrahlung cross section as described in \citet{2008ApJ...674..570E} and \citet{2011A&A...536A..93J} and using the cross-section shown in \citet{1953PhRv...90.1030G} and \citet{1972SoPh...25..425H} given by

\begin{equation}
\sigma_{I}(E,\epsilon,\Theta)=\sigma_{\perp}(E,\epsilon,\Theta)+\sigma_{\parallel}(E,\epsilon,\Theta),
\end{equation}
\begin{equation}
\sigma_{Q}(E,\epsilon,\Theta)=(\sigma_{\perp}(E,\epsilon,\Theta)-\sigma_{\parallel}(E,\epsilon,\Theta))\cos2\Theta ,
\end{equation}
and
\begin{equation}
\sigma_{U}(E,\epsilon,\Theta)=(\sigma_{\perp}(E,\epsilon,\Theta)-\sigma_{\parallel}(E,\epsilon,\Theta))\sin2\Theta,
\end{equation}
where $\epsilon$ is the X-ray photon energy and $\sigma_{\perp}(E,\epsilon,\Theta)$ and $\sigma_{\parallel}(E,\epsilon,\Theta)$ are the perpendicular and parallel components of the
bremsstrahlung cross-section. Subscripts $I, Q, U$ denote the cross section used for the total X-ray flux ($I$) and linear polarization components ($Q,U$) respectively, and

\begin{equation}
\cos\Theta=\cos\theta\cos\beta+\sin\theta\sin\beta\cos\Phi,
\label{eq:cosTheta}
\end{equation}

relates $\theta$ the photon emission angle measured from the local solar vertical, $\beta$ the electron pitch-angle (e.g. the angle between the electron velocity and the magnetic field) and $\Phi$ the electron azimuthal angle measured in the plane perpendicular to the local solar vertical.

Using the above cross sections the resulting photon flux $I$ and each specific linear polarization state $Q$ and $U$ can be written as:

\begin{equation}
I(\epsilon,\theta)\propto\int^{\infty}_{E=\epsilon}\int^{2\pi}_{\Phi=0}\int^{\pi}_{\beta=0}F(E,\beta)
\sigma_{I}(E,\epsilon,\Theta) \sin\beta d\beta d\Phi dE
\label{eq:I}
\end{equation}

\begin{equation}
Q(\epsilon,\theta)\propto\int^{\infty}_{E=\epsilon}\int^{2\pi}_{\Phi=0}\int^{\pi}_{\beta=0}F(E,\beta)\sigma_{Q}(E,\epsilon,\Theta)
 \sin\beta d\beta d\Phi dE,
\label{eq:Q}
\end{equation}

\begin{equation}
U(\epsilon,\theta)\propto\int^{\infty}_{E=\epsilon}\int^{2\pi}_{\Phi=0}\int^{\pi}_{\beta=0}F(E,\beta)\sigma_{U}(E,\epsilon,\Theta)
\sin\beta d\beta d\Phi dE.
\label{eq:U}
\end{equation}

\subsection{X-ray transport effects and determining the polarization observables}
Depending on the directivity, some fraction of the emitted X-rays are transported to the photosphere. Here Compton backscattering will change the properties of these X-rays before their escape towards the observer. This is known as the X-ray albedo component \citep[e.g.][]{1972ApJ...171..377T,1973SoPh...29..143S,1978ApJ...219..705B} and all X-ray observables including the flux and polarization spectra are altered by this albedo component. Hence, we employ the code of \citet{2011A&A...536A..93J} to create the X-ray albedo component for all simulation runs. A full discussion regarding this code can be found in \citet{2011A&A...536A..93J}.

Once the X-ray albedo component has been included, two polarization observables, the degree of polarization (DOP) and polarization angle $\Psi$ can be calculated using

\begin{equation}
DOP=\frac{\sqrt{Q^{2}+U^2}}{I},
\label{eq:DOP}
\end{equation}
and
\begin{equation}
\Psi=\frac{1}{2}\arctan\left(\frac{-U}{-Q}\right).
\label{eq:Psi}
\end{equation}

The negatives in Equation \ref{eq:Psi} ensure that $\Psi=0^{\circ}$ corresponds to a dominant polarization direction parallel to the local solar radial direction (negative DOP) and $\Psi=90^{\circ}$ corresponds to a dominant polarization direction perpendicular to the local solar radial direction (positive DOP). We note that high energy electrons (MeV) can produce spatially integrated $\Psi=90^{\circ}$ and thus positive DOP, due to electrons with higher energies scattering through larger angles, providing a useful diagnostic for the presence of MeV electrons. Moreover, \citet{2008ApJ...674..570E} showed that values of spatially integrated $\Psi$ other than $0^{\circ}$ or $90^{\circ}$ are possible when the tilt of the flare loop moves away from the local vertical direction (loop tilt $\tau$). Finally, we note that in the case of spatially resolved observations, the angle of polarization $\Psi$ can have values other than $\Psi=0^{\circ}$ or $\Psi=90^{\circ}$ due to the Compton scattered albedo component as shown in \citet{2011A&A...536A..93J}, providing a detailed probe of electron directivity in the chromosphere.

Here we only study spatially integrated X-ray data for relatively low electron energies $<300$~keV and for flare loops aligned along the local solar vertical. Hence $\Psi=0^{\circ}$ and DOP is negative for all simulations. Therefore, since $\Psi$ and the sign of DOP do not change, we only show the DOP results (and as a percentage only). 

\begin{figure*}
\centering
\includegraphics[width=0.59\textwidth]{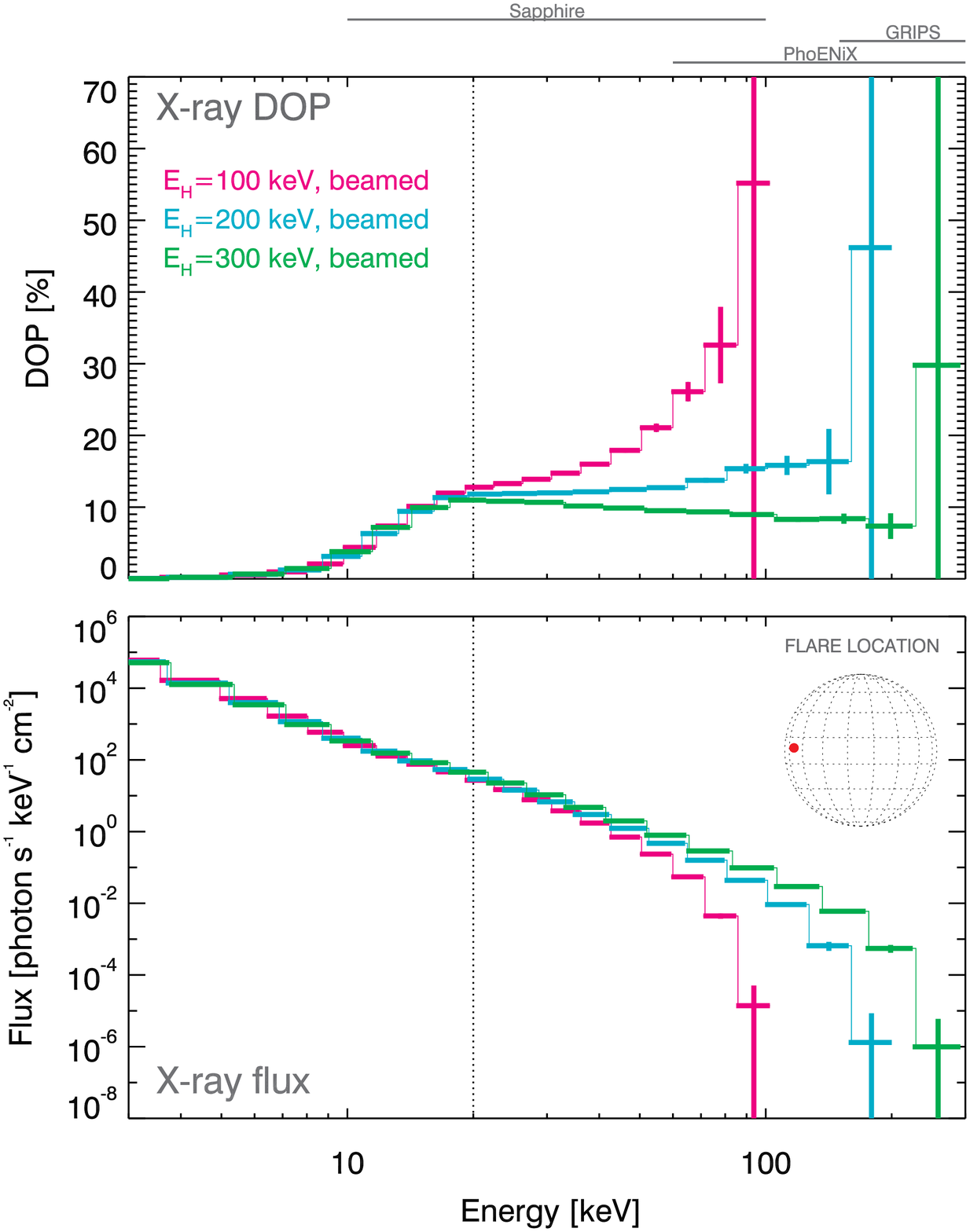}
\caption{Resulting spatially integrated X-ray DOP and flux spectra for injected electron distributions with different high energy cutoffs of $E_{H}=100$~keV, $E_{H}=200$~keV and $E_{H}=300$~keV. Each use the following identical electron and plasma properties of: $\delta=5$, $E_{c}=20$~keV (vertical grey dotted line), a beamed distribution and $\dot{N}=7\times10^{35}$~e s$^{-1}$, and coronal plasma properties of: $n=3\times10^{10}$~cm$^{-3}$ and $T=20$~MK, plotted for a flare located at a heliocentric angle of $60^{\circ}$. All DOP spectra include an albedo component and a coronal thermal component ($EM=0.9\times10^{48}$~cm$^{-3}$). Higher $E_{H}$ produces a clear flattening in the DOP spectra after the low energy cutoff. No turbulent scattering is present.}  \label{fig2}
\end{figure*}

\section{Prospective X-ray polarization instrumentation}\label{instruments}

Instrumentation to specifically detect solar flare HXR polarization are currently being researched and developed and three such instruments are summarised here.
Currently, the most advanced concept is the Gamma-Ray Imager/Polarimeter for Solar flares (GRIPS, \citet{2016SPIE.9905E..2QD}).
GRIPS is a balloon-borne telescope designed to study solar-flare particle acceleration and transport. 
It has already flown once in Antarctica in January 2016, and will be re-proposed for flight during the next Solar Maximum.
GRIPS can do imaging spectro-polarimetry of solar flares in the $\sim$150  keV to $\sim$10 MeV range, with a spectral resolution of a few keV, and an angular resolution of $\sim12.5\arcsec$. 
GRIPS's key technological improvements over the current solar state of the art at HXR/gamma-ray energies, RHESSI, include 3D position-sensitive germanium detectors (3D-GeDs) and a single-grid modulation collimator, the multi-pitch rotating modulator (MPRM). 
Focusing optics or Compton imaging techniques are not adequate for separating magnetic loop footpoint emissions in flares over the GRIPS energy band, and indirect imaging methods must be employed.
The GRIPS MPRM covers 13 spatial scales from $12.5\arcsec$ to $162\arcsec$. 
For comparison, RHESSI could only image gamma-ray emissions at two spatial scales ($35\arcsec$ and $183\arcsec$).
For photons that Compton scatter, usually $\gtrsim 150$~keV, the energy deposition sites can be tracked, providing polarization measurements as well as enhanced background reduction through Compton imaging.  
The nominal GRIPS balloon payload has a minimum detectable polarization (MDP) signal of $\sim$3\% in the 150-650 keV band for 2002-July-23 X-flare, 
while a spacecraft version will likely be closer to $\sim$1\%.
While we plan to discuss the spatially resolved observations of GRIPS in another study, here we concentrate on the usefulness of spatially integrated observations of the polarization (DOP) spectrum at lower energies. 

In the HXR ($\sim$10-100 keV) regime, the proposed Sapphire (Solar Polarimeter for Hard X-rays, \citet{2019AGUFMSH31C3310S}) concept aims to do spatially-integrated spectro-polarimetry of solar flares from CubeSat platforms. A single Sapphire module is expected to be able to detected polarization in the $\sim$1.5\% threshold ($\ge20$ keV, 3-sigma) for a similar flare. Sapphire modules are designed to be stackable, with the decrease (improvement) in MDP for $N$ modules roughly behaving as $1/\sqrt{N}$.

Other proposed missions include the Japanese PhoENiX mission (Physics of Energetic and Non-thermal plasmas in the X (= magnetic reconnection) region, \citet{2019AAS...23412603N}) which will have an X-ray spectro-polarimeter on board measuring over an energy range of 60-300~keV (and 20-300 keV for spectroscopy). 

\section{Results}\label{results}

We investigate how spatially integrated X-ray DOP changes with anisotropy, high energy cutoff and non-collisional turbulent scattering. DOP (and indeed polarization angle) will vary with the flare geometrical properties such as heliocentric angle and the properties of the flare loop such as loop tilt to the local vertical direction, but these properties can always be estimated from flare imaging (see Appendix \ref{app_A1}, Figure \ref{figA1}). In the following sections, all results are shown for a heliocentric angle of $60^{\circ}$ and a loop tilt of $0^{\circ}$ (the loop apex is parallel to the local vertical direction).

\begin{figure*}
\centering
\includegraphics[width=0.59\textwidth]{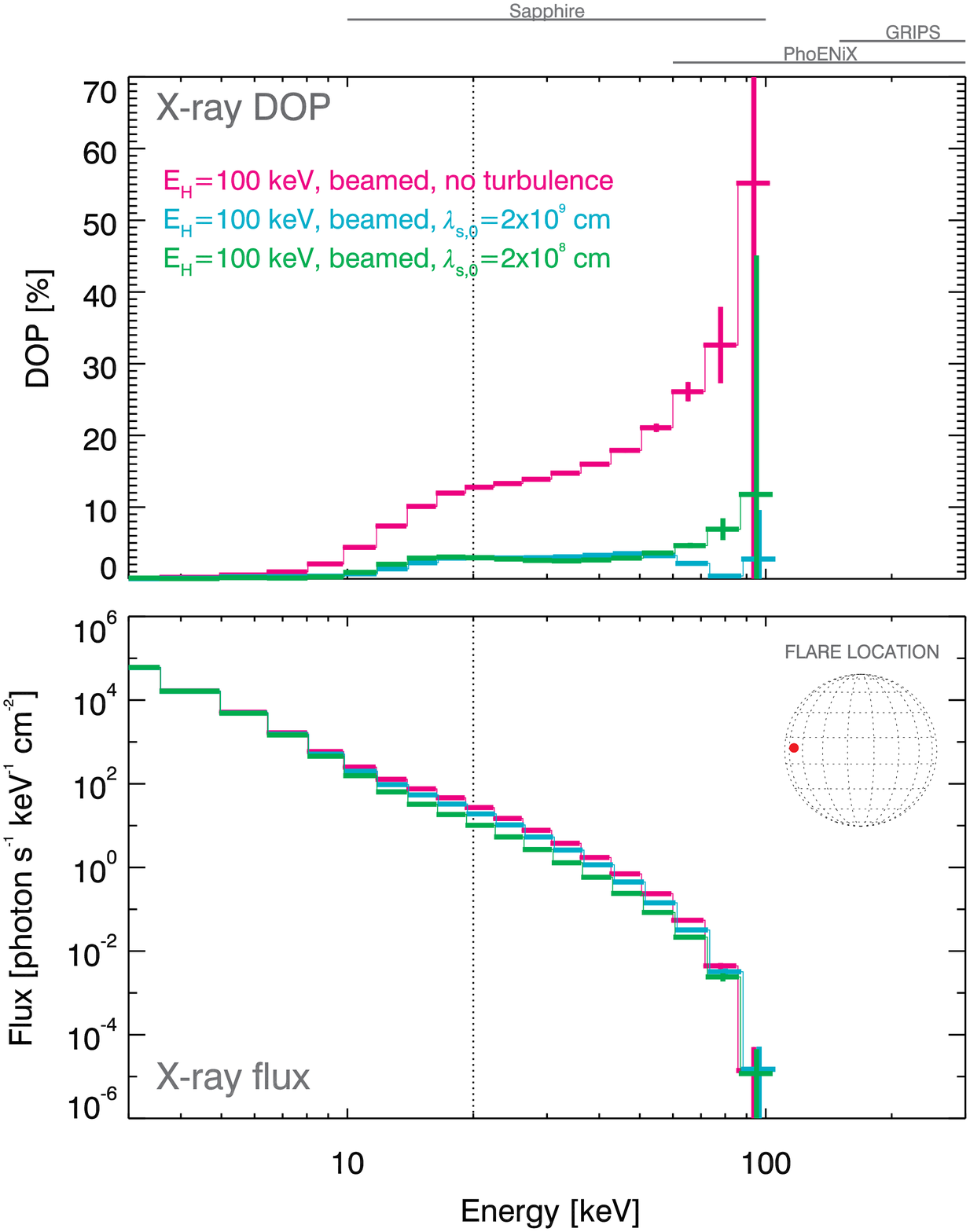}
\caption{Resulting X-ray DOP and flux spectra for injected beamed electron distributions with $E_{H}=100$~keV without turbulent scattering, and with turbulent scattering (using $\lambda_{s, 0}=2\times10^{8}$ cm or $\lambda_{s, 0}=2\times10^{9}$ cm) situated in the coronal loop over a distance of [-10\arcsec,+10\arcsec] from the loop apex. Each use the following identical electron and plasma properties of: $\delta=5$, $E_{c}=20$~keV (vertical grey dotted line), a beamed distribution and $\dot{N}=7\times10^{35}$~e s$^{-1}$, and coronal plasma properties of: $n=3\times10^{10}$~cm$^{-3}$ and $T=20$~MK, plotted for a flare located at a heliocentric angle of $\theta=60^{\circ}$. All DOP spectra include an albedo component and a coronal thermal component ($EM=0.9\times10^{48}$~cm$^{-3}$). The results suggest that a lack of coronal turbulent scattering could be detectable from the DOP spectrum.}  \label{fig3}
\end{figure*}

\subsection{X-ray polarization and electron anisotropy}\label{anisotropy}

Traditionally, the prime diagnostic of X-ray polarization is its direct link with electron directivity.  Figure \ref{fig1} shows the spatially integrated DOP (\%) versus X-ray energy for four different injected electron anisotropies from completely beamed to isotropic\footnote{For comparison, the emitting electron distribution has been forced to be completely isotropic and the DOP spectrum peaking at 2-3\% results purely from the backscattered albedo component \citep[e.g.][]{1978ApJ...219..705B,2011A&A...536A..93J}.}, for a given set of otherwise identical coronal plasma parameters and electron spectral and spatial parameters. The resulting DOP spectra contains an albedo component and a thermal component as described in Section \ref{model}.  As expected, the DOP at all energies clearly decreases with increasing injected electron isotropy. In Figure \ref{fig1}, we plot the results for an average coronal number density of $n=3\times10^{10}$~cm$^{-3}$ and coronal temperature of $T=20$~MK. The plasma properties of the corona do change the resulting DOP slightly, particularly at lower energies below 50 keV, however, properties such as the average coronal number density can be estimated from the X-ray flux spectrum and/or EUV spectral observations, and applied to the DOP spectrum. 

In Figure \ref{fig1}, we plot the X-ray flux and DOP over an energy range of $\approx3-100$~keV, where the flare count rates are highest (most flare spectra are steeply decreasing power laws). As an example in Figures \ref{fig1}-\ref{fig3}, we calculate sensible flux and DOP error values using ${\rm counts} = {\rm flux} \times {\rm effective\;area}\;(A) \times {\rm time\;bin}\;(\Delta T) \times {\rm energy\;bin}\;(\Delta E)$ and assuming that ${\rm photons} = {\rm counts}$. The error on the flux is then calculated as ${\rm flux\;error} = \frac{\sqrt{{\rm counts}}}{A \Delta t \Delta E}$ and the corresponding DOP error as ${\rm DOP\;error}={\rm flux\;error}\times\frac{{\rm DOP}}{{\rm flux}}$. Here we use $A=5$~cm$^{2}$ and $\Delta t=120$~s, with $\Delta E$ shown in each figure. We plot the resulting DOP spectra for a flare located at a heliocentric angle of 60$^{\circ}$ (the DOP should grow as the heliocentric angle approaches the limb).

\subsection{X-ray polarization and the high energy cutoff}\label{he_cutoff}
Another important diagnostic of the acceleration mechanism is the high energy cutoff (the highest energy electrons produced by the acceleration process). The high energy cutoff is also dependent on the spatial extent of the acceleration region and the acceleration time, making it an important modelling constraint. Although the presence of high (MeV) energy electrons might be determinable from microwave observations if present e.g. \citet{2002ApJ...580L.185M} and \citet{2018ApJ...863...83G}, we show that the high energy cutoff changes the trend in the DOP spectra (and also the sign of the polarization angle $\Psi$ for MeV energies as discussed in \citet{2011A&A...536A..93J}). We show that the presence of higher energy keV electrons in the distribution can decrease the overall DOP at all observed energies.

In Figure \ref{fig2}, we plot the resulting DOP spectra from three injected electron distributions with different high energy cutoffs of 100 keV, 200 keV and 300 keV. After 20~keV, we can see that the gradient of the DOP spectrum decreases with an increase in the high energy cutoff $E_{H}$, producing a flattening and then a decrease in DOP with energy as the high energy cutoff increases to $E_{H}=300$~keV, over a typical observed flare energy range of $\approx10-100$~keV. 

This occurs because the resulting X-ray directivity is also dependent on the bremsstrahlung cross section. For example, an 80 keV X-ray is more likely to be emitted by an electron of energy 100 keV than an electron of energy 80 keV (i.e. magnitude of the cross section), with the directivity of the emitted 80 keV X-ray decreased (i.e. the cross section is more isotropic; see Appendix \ref{app_B1}, Figure \ref{figA3}). However, solar flare electron energy spectra are steeply decreasing power laws and to check that the X-ray emission from high energy electrons can indeed contribute enough emission to significantly decrease the X-ray directivity and DOP, two electron distributions with $E_{H}=100$ keV and $E_{H}=300$ keV, and $\delta=5$ are compared (see Appendix \ref{app_B1}, Figure \ref{figA2}). By studying the emission from this electron distribution above and below an example energy of 40 keV separately, we see that for X-rays above 25-30 keV, the contribution from higher energy electrons dominates, with this contribution dominating more at lower X-ray energies for higher $E_{H}$.

This ultimately means that electron distributions with higher electron energies will produce lower energy X-rays with a smaller DOP, even for the same injected beaming. This is an important diagnostic tool that can help to constrain the highest energies in the electron distribution, even when they are completely undetectable by other means. The X-ray DOP spectrum is always a result of both the electron anisotropy and high energy cutoff and both parameters should be determined in tandem.

\begin{figure*}[t]
\centering
\includegraphics[width=0.49\textwidth]{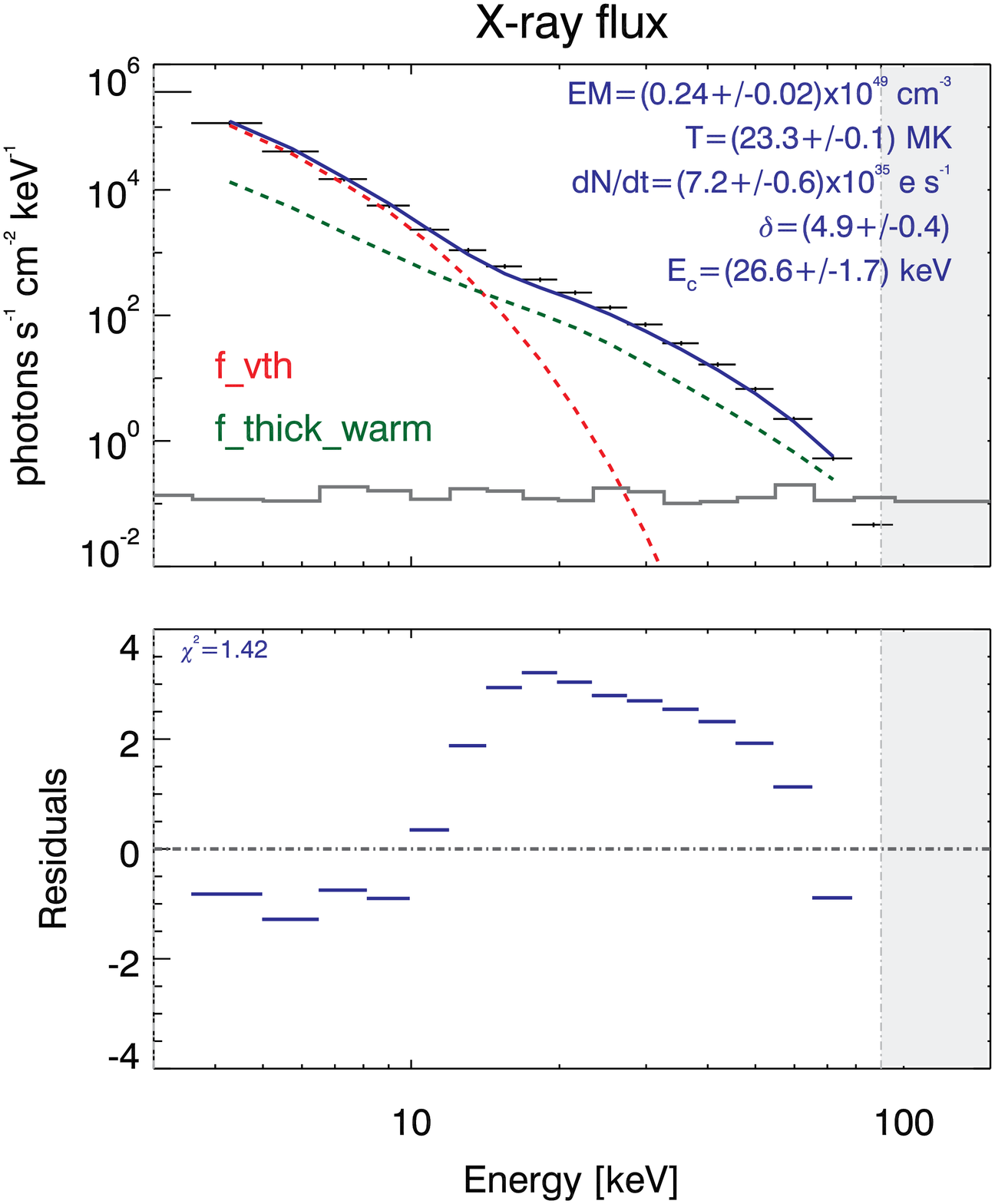}
\includegraphics[width=0.49\textwidth]{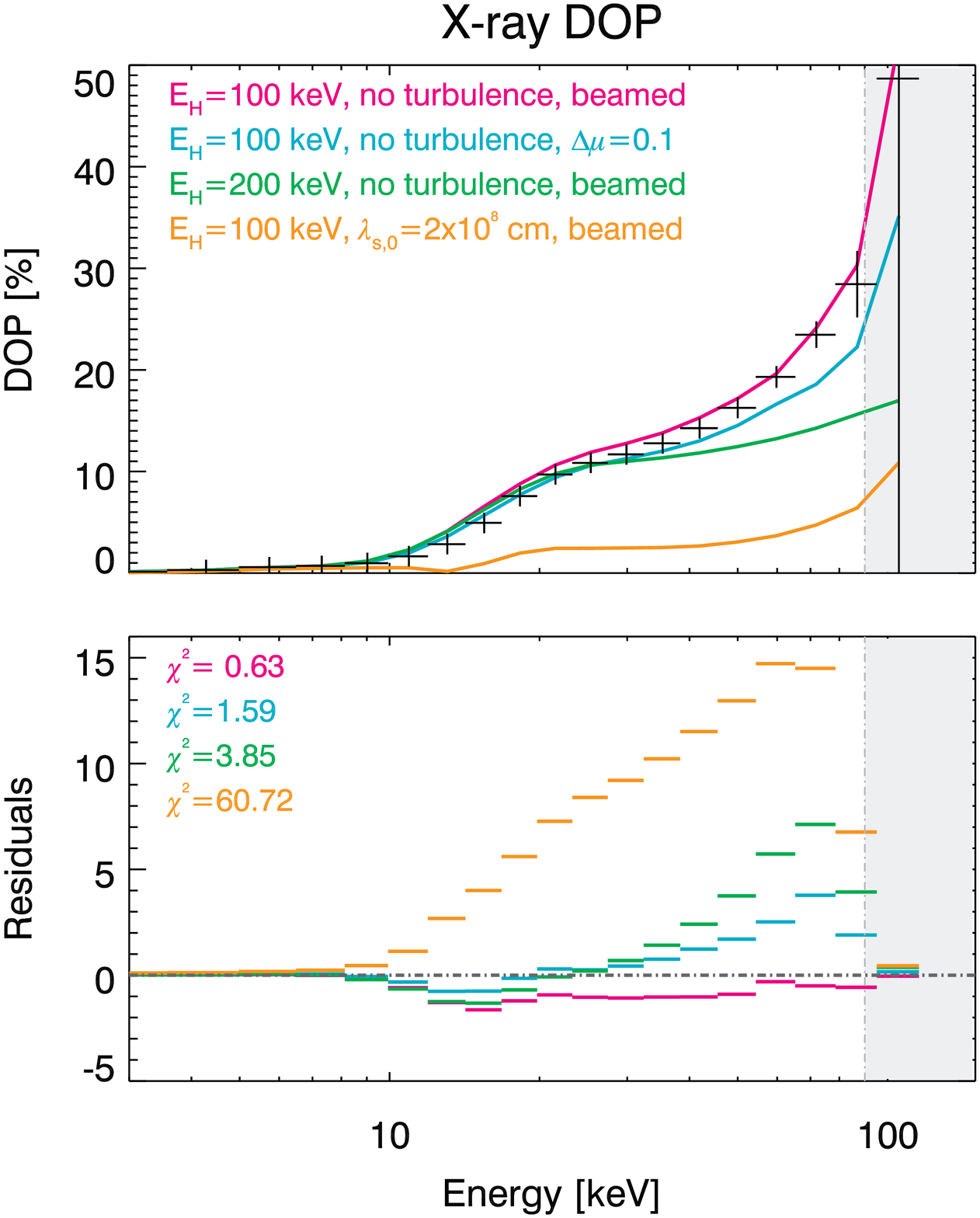}
\caption{Example of how the X-ray flux and DOP spectra can be used together to estimate the properties of the accelerated electron distribution. Applying the warm target function (f\_thick\_warm in OSPEX) to the X-ray flux helps to constrain the coronal plasma properties and the following acceleration parameters: low energy cutoff $E_{c}$, the spectral index $\delta$ and the rate of acceleration $\dot{N}$. Ignoring any other transport mechanisms, and once constrained, the resulting X-ray DOP spectra should only be dependent upon the high energy cutoff $E_{H}$ and acceleration anisotropy $\Delta\mu$, and turbulent scattering (using $\lambda_{s,\,0}$), if present. Top panels:  (left) The `flux data' (black) and the resulting f\_vth+f\_thick\_warm fits to the X-ray flux spectrum (red and green respectively, this also contains an albedo component not shown), and (right) the `DOP data' (black) and four simulation runs using the constrained properties from the X-ray flux spectra and the values of $E_{H}$, $\Delta\mu$ and $\lambda_{s,\,0}$ shown. Bottom panels: Residuals and resulting reduced $\chi^{2}$ for each fit. Both the X-ray flux and DOP are fitted between 3 and 90 keV (grey dash-dot line).}
\label{fig4a}
\end{figure*}

\begin{table*}
\begin{center}
\normalsize
\begin{tabular}{|c|c|c|c|}
\hline
Parameter & Simulation & Estimated & Method\\
\hline
$EM$ &$0.3\times10^{49}$ cm$^{-3}$&$(0.24\pm0.02)\times10^{49}$ cm$^{-3}$&X-ray flux, f\_vth\\
$T$ & $25$ MK& $23.3\pm0.1$ MK&X-ray flux, f\_vth\\
$n$ & $6\times10^{10} $cm$^{-3}$&$(4.9\pm0.4)\times10^{10}$ cm$^{-3}$&X-ray flux, f\_vth ($EM$) and $V$\\
$L$ & $24\arcsec$&$24\arcsec$&X-ray imaging\\
$\dot{N}$ &$7\times10^{35}$ e s$^{-1}$ &$(7.19\pm0.58)\times10^{35}$ e s$^{-1}$&X-ray flux, f\_thick\_warm\\
$\delta$ & $4.6$&4.93$\pm$0.39&X-ray flux, f\_thick\_warm\\
$E_{c}$ & $24.0$ keV&26.6$\pm$1.7 keV&X-ray flux, f\_thick\_warm\\
$E_{H}$ & $113$ keV&$\approx 100$ keV&X-ray DOP\\
$\Delta\mu$ & Beamed&Beamed ($\Delta\mu<0.1$)&X-ray DOP\\
$\lambda_{s,0}$ &No turbulence&No turbulence&X-ray DOP\\
\hline
\end{tabular}
\caption{Comparison of the determined parameters and those input into the simulation. We demonstrate that using the X-ray flux and DOP spectra in tandem can help us to estimate all of the acceleration parameters together and help to constrain the acceleration processes in flares.}
\label{t1a}
\end{center}
\end{table*}

\subsection{Transport versus injection properties}\label{transport}

Other transport effects such as turbulent scattering also alter the properties of the electron distribution, including the anisotropy and hence the X-ray polarization. An estimation of the plasma properties from the X-ray flux spectrum, imaging and EUV observations can determine the effects due to collisions, however, constraining the presence and properties of additional non-collisional transport effects outside of the acceleration region can be challenging in the majority of flares. Since DOP is sensitive to changes in anisotropy, we test if the DOP spectra can be used to separate acceleration isotropy with isotropy produced by additional transport processes in the corona outside of the acceleration region.

In Figure \ref{fig3}, as one example, we plot the DOP versus energy for three identical beamed electron distributions with $E_{H}=100$~keV, with no coronal turbulence and with turbulence ($\lambda_{s,\,0}=2\times10^{8}$ cm and $\lambda_{s,\,0}=2\times10^{9}$ cm) situated in the coronal loop over a distance of [-10\arcsec,+10\arcsec] from the loop apex. As expected, the presence of turbulence in the corona increases the electron isotropy and hence, reduces the DOP at all energies. Even very beamed distributions produce low levels of DOP when turbulent scattering is present, as shown in Figure \ref{fig3}.

Using DOP only, it is difficult to distinguish between an initially isotropic electron distribution and the presence of strong turbulence in the corona outside of the acceleration region, although it is suggestive that if strong turbulence is present in the corona then it is also highly likely to be present in the coronal acceleration region. Turbulence greatly affects higher energy electrons and isotropises them quickly leading to low DOP at all energies. The results indicate that a lack of turbulence in the flaring corona can be determined from the DOP spectrum. Figure \ref{fig3} also shows interesting differences between strong and milder turbulent scattering (i.e. slightly greater DOP at large energies from stronger turbulent scattering) that will be investigated further for different turbulent conditions beyond the scope of this study.

\subsection{Extracting parameters from a combined X-ray flux and DOP spectral analysis}\label{find_dop}

The above results show that X-ray polarization is dependent on several important electron acceleration properties and the coronal plasma properties. Moreover, this shows that the DOP spectrum is a powerful diagnostic tool, particularly when used alongside the X-ray flux spectrum.

As a preliminary demonstration of how we could extract vital acceleration parameters from combined X-ray flux and DOP spectra together, we simulate a flare at a heliocentric angle of $60^{\circ}$ with certain accelerated electron and background coronal properties (see Table \ref{t1a}), producing the resulting X-ray flux and DOP spectra shown in Figure \ref{fig4a}. To each spectrum, we add noise and we assume a spacecraft background level of $10^{-2}$ photons s$^{-1}$ cm$^{-2}$ keV$^{-1}$ at all energies. We now treat the outputs as observational data with unknown properties. Firstly, to the X-ray flux spectrum, we apply the Solar Software (SSW)/OSPEX fitting function routines (as with real flare data). We fitted the X-ray flux spectrum with an isothermal function (f\_vth in OSPEX) and warm-target fitting function (f\_thick\_warm in OSPEX), using the steps described in \citet{2019ApJ...871..225K}. In this fit, as is usual with data, we set $E_H=200$~keV, a value above the highest energy used in the fitting (90 keV). From fitting, we determine the following accelerated electron parameters: $\dot{N}=(7.19\pm0.58)\times10^{35}$~e s$^{-1}$, $\delta=-4.93\pm0.39$, $E_{c}=26.6\pm1.7$~keV, and the surrounding coronal plasma parameters of $T=23.3\pm0.1$~MK, $EM=(0.24\pm0.02)\times10^{49}$~cm$^{-3}$ and $n=\sqrt{EM/V}=(4.9\pm0.4)\times10^{10}$~cm$^{-3}$ where $V$ is the volume of the hot plasma, assuming a sensible flare volume of $V\approx1\times10^{27}$~cm$^{-3}$ and loop length $L=24\arcsec$, as shown in Table \ref{t1a}. Once these parameters are constrained, we can fix them in the simulation. Then, the only unknowns are $E_{H}$, $\Delta\mu$ and any coronal loop turbulence described here using $\lambda_{s\,,0}$. Using different simulation runs, we can constrain and determine these parameters by comparison with the observed flare X-ray DOP spectrum. In Figure \ref{fig4a}, four such runs are shown and compared with the DOP spectrum; the runs use different anisotropies, high energy cutoffs and one with turbulent scattering filling the entire corona loop. The residuals and goodness of fit $\chi^{2}$ (calculated as the sum of the residuals divided by the number of fitted energy bins) of each resulting simulation `model' are also shown as an example of how observations and models can be compared.

In Table \ref{t1a}, we show all the determined electron and plasma parameters, the method used to obtain each parameter and compare with the actual parameters that were used in the original `data' simulation. Using X-ray flux and X-ray DOP observations together provides us with a fuller understanding of the solar flare acceleration mechanism. Our analysis shows that current modelling is capable of producing estimates of $E_{H}$ and $\Delta\mu$, and determining whether turbulence exists in the corona. It will be possible to determine the uncertainties on these variables using many simulation runs and a full Monte Carlo parameter space analysis. However, this full uncertainty analysis is beyond the scope of the current work, since the aim here is to demonstrate how the DOP spectrum can be used as a powerful diagnostic tool alongside the X-ray flux spectrum if the data becomes available.

The DOP is also sensitive to other electron acceleration parameters such as e.g. spectral index and breaks in the spectrum. However, as shown these parameters can be constrained from the X-ray flux spectrum before analysing the DOP spectrum, showing the importance of studying the X-ray flux and DOP spectra in tandem.

\section{Summary}\label{conclusions}
X-ray polarimetry is a vital tool for constraining the solar flare acceleration mechanism especially when used alongside the X-ray flux spectrum. The X-ray DOP spectrum is highly sensitive to currently unknown properties such as the accelerated electron pitch-angle distribution, highest energy accelerated electrons and the presence of turbulence in the corona. We have simulation tools available to analyse the X-ray DOP spectrum in detail, if the observations become available and although imaging polarimetry is highly desired, our results show that missions without imaging can also provide strong constraints on electron anisotropy and the high energy cutoff.

In this paper we specifically discussed transport processes that can change the pitch angle distribution of electrons such as Coulomb collisions and turbulent scattering due to magnetic fluctuations but other transport processes may be present. We used a model with a simple homogenous background plasma in the corona since we cannot study the individual plasma properties of each flare. Spatial variations in parameters such as number density along the loop will cause changes in the DOP spectrum. If such changes can be inferred from the data, then they can be incorporated into the model before the inference of acceleration properties.

\begin{acknowledgements}
The development of the electron and photon transport models are part of an international team grant (``Solar flare acceleration signatures and their connection to solar energetic particles'' \url{http://www.issibern.ch/teams/solflareconnectsolenerg/}) from ISSI Bern, Switzerland. NLSJ acknowledges IDL support provided by the UK Science and Technology Facilities Council (STFC).  We thank the anonymous referee for their useful comments that helped to improve the text.
\end{acknowledgements}


\begin{appendix}

\section{Change in DOP versus energy with heliocentric angle and loop tilt}\label{app_A1}

\begin{figure*}[hbtp!]
\centering
\includegraphics[width=0.99\linewidth]{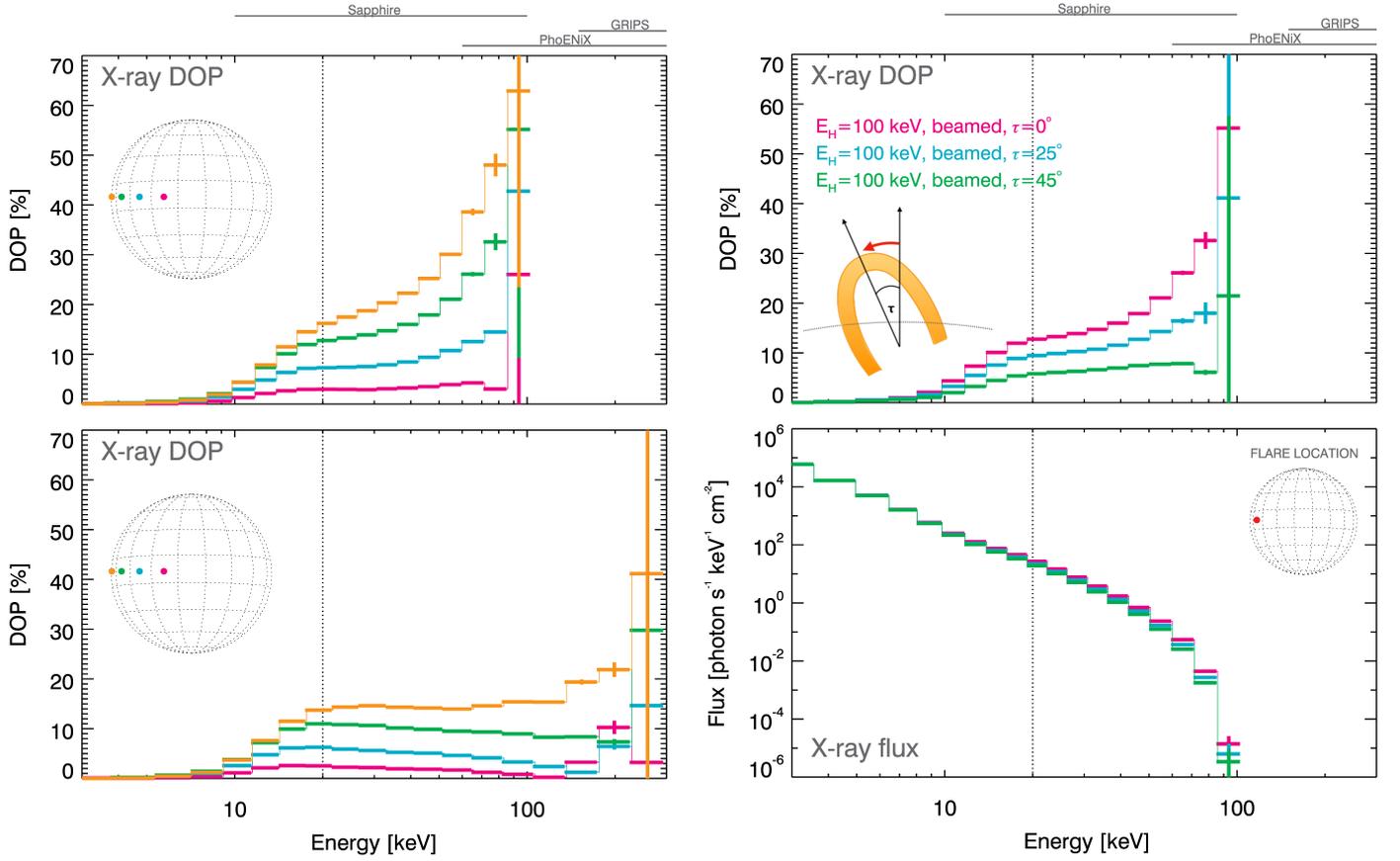}
\caption{Left: Change in spatially integrated X-ray DOP versus energy for four different heliocentric angles of $20^{\circ}$, $40^{\circ}$, $60^{\circ}$ and $80^{\circ}$ (as shown by the coloured dots). In this example, all electron distributions are beamed with $E_{H}=100$~keV (top panel) and $E_{H}=300$~keV (bottom panel). Right: Change in spatially integrated X-ray DOP and flux versus energy for different loop tilts of $\tau=0^{\circ}$, $25^{\circ}$, $45^{\circ}$ (the apex of the loop is tilted by an angle relative to the local vertical - see small cartoon). Each (left and right) uses the following identical electron properties of: $\delta=5$, $E_{c}=20$~keV (grey dotted line), and $\dot{N}=7\times10^{35}$~e $s^{-1}$, and corona plasma properties of: $n=3\times10^{10}$~cm$^{-3}$ and $T=20$~MK. All spectra include an albedo component and a coronal background thermal component ($EM=0.9\times10^{48}$~cm$^{-3}$).}
\label{figA1}
\end{figure*}

The spatially integrated DOP changes with both flare location on the solar disk (heliocentric angle) and with loop tilt (how the loop apex is tilted with respect to the local vertical; the polarization angle also changes with loop tilt, \citet{2008ApJ...674..570E}). However, both the heliocentric angle and loop tilt can be estimated from imaging the flare in different wavelengths (e.g. EUV). We show some examples of how DOP changes with heliocentric angle and loop tilt in Figure \ref{figA1}.

\section{Bremsstrahlung cross section and the contribution from high energy electrons}\label{app_B1}

\begin{figure*}
\centering
\includegraphics[width=0.49\textwidth]{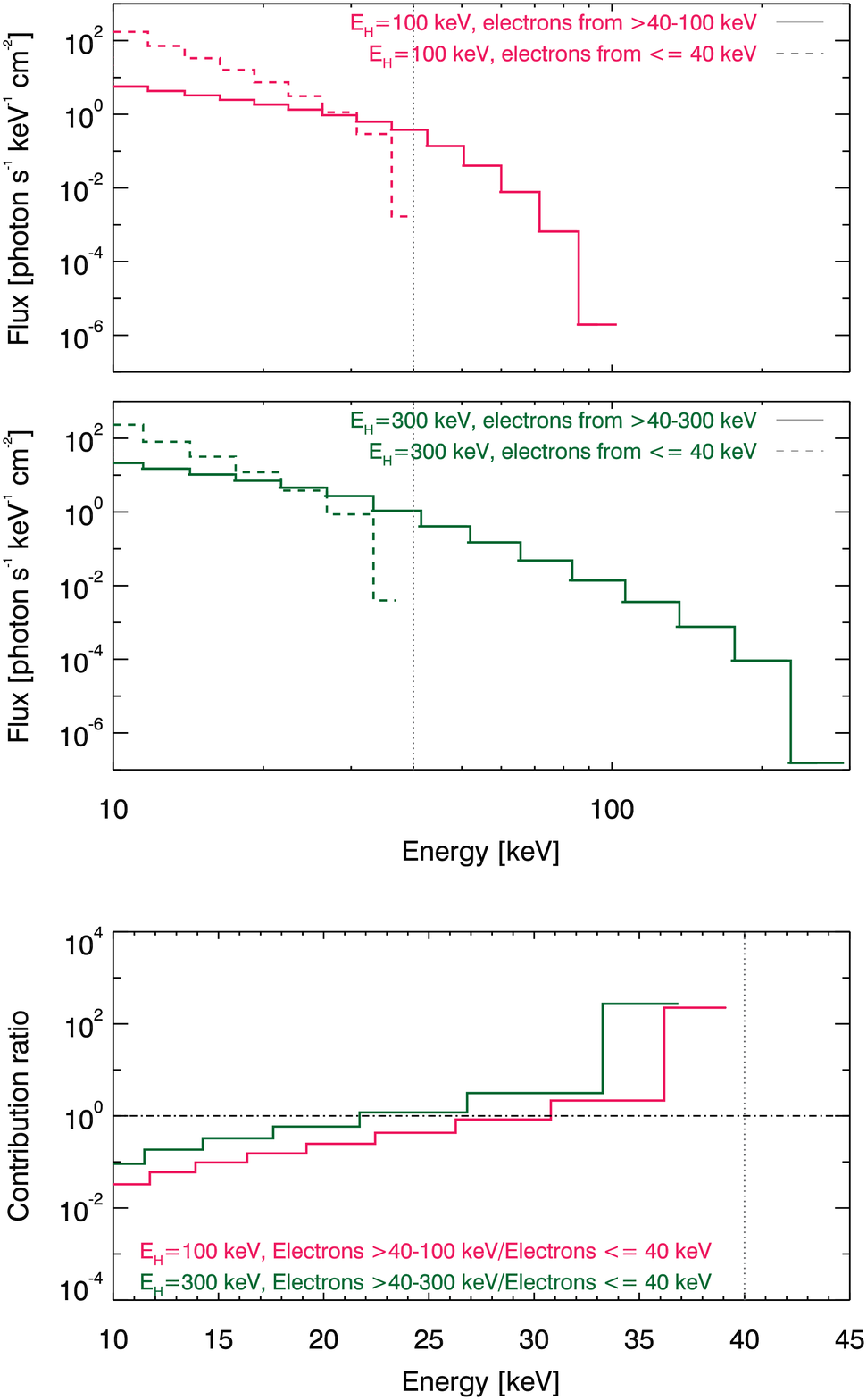}
\caption{Resulting X-ray fluxes from electrons of energy $E\le40$ keV (dashed line) and from electrons of energy $E>40$~keV (solid line) separately, for two different high energy cutoff of $E_{H}=100$~keV (top) and $E_{H}=300$~keV (middle). Bottom: X-ray contribution ratio (the X-ray emission from $E>40$~keV divided by the X-ray emission from $E\le40$ keV) versus energy up to 40 keV (vertical grey dotted line), for electron distributions with $E_{H}=100$~keV and $E_{H}=300$~keV. The horizontal black dashed-dot line denotes a ratio of 1.}  \label{figA2}
\end{figure*}

In subsection \ref{he_cutoff}, we determined that a high energy cutoff leads to lower DOP across all energies and explained that this is due to the production of a more isotropic X-ray distribution when higher energy electrons are present. To confirm this, in Figure \ref{figA2} we plot the resulting X-ray fluxes from electrons of energy $E\le40$ keV (dashed line) and from electrons of energy $E>40$~keV (solid line) separately, but from the same emitting electron distribution. This is shown for an electron distribution with a high energy cutoff of $E_{H}=100$~keV (top panel) and $E_{H}=300$~keV (middle panel), and a spectral index of $\delta=5$. The bottom panel of Figure \ref{figA2} shows the X-ray contribution ratio (defined as the X-ray emission from $E>40$~keV divided by the X-ray emission from $E\le40$ keV) versus energy up to 40 keV, for electron distributions with $E_{H}=100$~keV and $E_{H}=300$~keV respectively.

\begin{figure*}[hbtp!]
\centering
\includegraphics[width=0.49\linewidth]{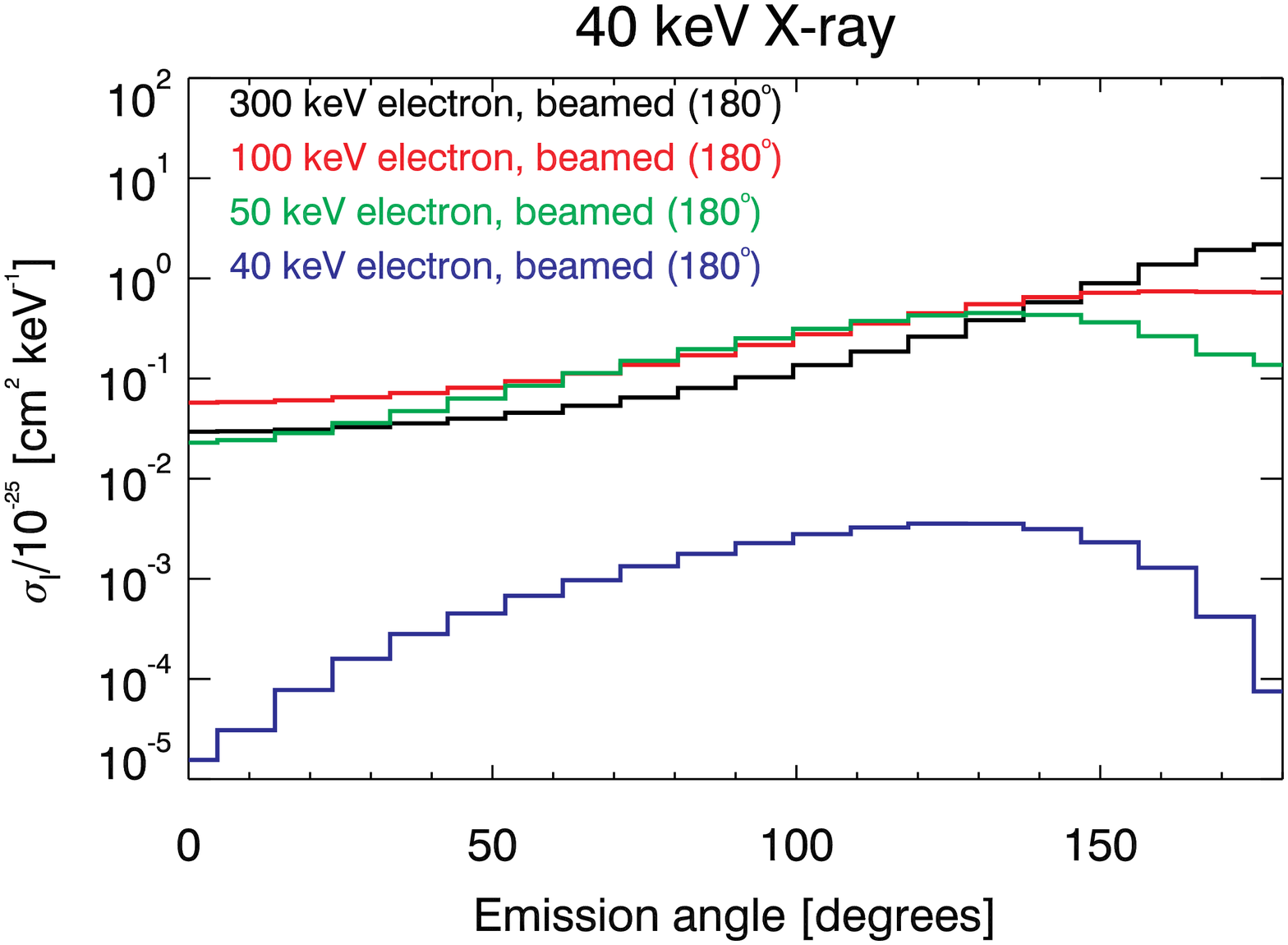}
\includegraphics[width=0.49\linewidth]{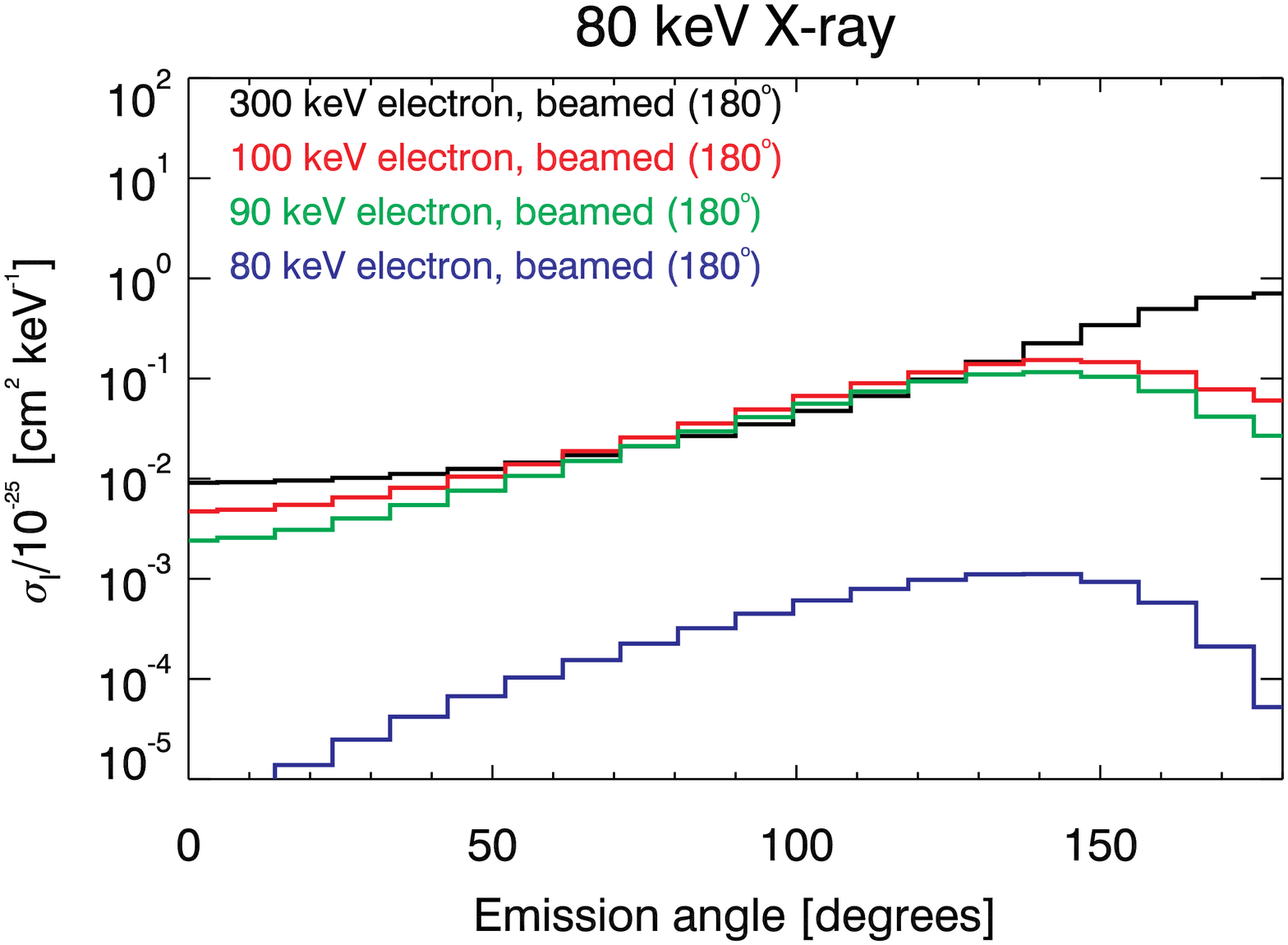}
\caption{Total bremsstrahlung cross section ($\sigma_{I}$) for the emission of a 40 keV X-ray by electrons of different of energies 40, 50, 100 and 300 keV (left) and for the emission of an 80 keV X-ray by electrons of different of energies 80, 90, 100 and 300 keV.}
\label{figA3}
\end{figure*}

Although, the contribution will vary with other properties such as spectral index $\delta$, we can see that for both $E_{H}=100$~keV and $E_{H}=300$~keV, the X-ray contribution from higher energy electrons ($E>40$~keV) dominates from above $25$ keV ($E_{H}=300$ keV) and above $30$ keV ($E_{H}=100$ keV). Therefore, it shows that the bulk of the X-rays $>25$ keV ($E_{H}=300$ keV) and  $>35$ keV ($E_H=100$ keV) will be dominated by a closer to isotropic X-ray distribution produced by electrons with $E>40$ keV (Figure \ref{figA3}).

\end{appendix}

\end{document}